\documentclass[reqno,11pt]{amsart}
\usepackage{amsmath, latexsym, amsfonts, amssymb, amsthm, amscd}
\usepackage{graphics,epsf,psfrag}
\numberwithin{equation}{section}

\newtheorem{theorem}{Theorem}[section]
\newtheorem{lemma}[theorem]{Lemma}
\newtheorem{proposition}[theorem]{Proposition}

\newtheorem{rem}[theorem]{Remark}

\newcommand{\ind}{\mathbf{1}}

\newcommand{\R}{\mathbb{R}}
\newcommand{\Z}{\mathbb{Z}}
\newcommand{\N}{\mathbb{N}}

\newcommand{\cN}{{\ensuremath{\mathcal N}} }

\newcommand{\bP}{{\ensuremath{\mathbf P}} }
\newcommand{\bE}{{\ensuremath{\mathbf E}} }

\DeclareMathSymbol{\leqslant}{\mathalpha}{AMSa}{"36} 
\DeclareMathSymbol{\geqslant}{\mathalpha}{AMSa}{"3E} 
\DeclareMathSymbol{\eset}{\mathalpha}{AMSb}{"3F}     
\newcommand{\dd}{\,\text{\rm d}}             

\newcommand{\bbE}{{\ensuremath{\mathbb E}} }

\newcommand{\bbP}{{\ensuremath{\mathbb P}} }

\newcommand{\bbS}{{\ensuremath{\mathbb S}} }

\newcommand{\ga}{\alpha}
\newcommand{\gb}{\beta}
\newcommand{\gd}{\delta}
\newcommand{\gep}{\varepsilon}       

\newcommand{\go}{\omega}

\newcommand{\gO}{\Omega}
\newcommand{\gl}{\lambda}

\makeatletter
\def\captionfont@{\footnotesize}
\def\captionheadfont@{\scshape}

\long\def\@makecaption#1#2{%
  \vspace{2mm}
  \setbox\@tempboxa\vbox{\color@setgroup
    \advance\hsize-6pc\noindent
    \captionfont@\captionheadfont@#1\@xp\@ifnotempty\@xp
        {\@cdr#2\@nil}{.\captionfont@\upshape\enspace#2}%
    \unskip\kern-6pc\par
    \global\setbox\@ne\lastbox\color@endgroup}%
  \ifhbox\@ne 
    \setbox\@ne\hbox{\unhbox\@ne\unskip\unskip\unpenalty\unkern}%
  \fi
  \ifdim\wd\@tempboxa=\z@ 
    \setbox\@ne\hbox to\columnwidth{\hss\kern-6pc\box\@ne\hss}%
  \else 
    \setbox\@ne\vbox{\unvbox\@tempboxa\parskip\z@skip
        \noindent\unhbox\@ne\advance\hsize-6pc\par}%
\fi
  \ifnum\@tempcnta<64 
    \addvspace\abovecaptionskip
    \moveright 3pc\box\@ne
  \else 
    \moveright 3pc\box\@ne
    \nobreak
    \vskip\belowcaptionskip
  \fi
\relax
}
\makeatother
\def\writefig#1 #2 #3 {\rlap{\kern #1 truecm
\raise #2 truecm \hbox{#3}}}

\newcommand{\tf}{\textsc{f}}
\newcommand{\tT}{\textsc{t}}
\newcommand{\tg}{\textsc{g}}

\newcommand{\Kbar}{\overline{K}}

\begin{document}

\title[Depinning and re-entrant transitions]{Force--induced depinning of  directed polymers}

\author{Giambattista Giacomin}
\address{Laboratoire de Probabilit{\'e}s de P 6\ \& 7 (CNRS U.M.R. 7599)
  and Universit{\'e} Paris 7 -- Denis Diderot, U.F.R.
  Math\'ematiques, Case 7012, 2 place Jussieu 75251 Paris cedex 05,
  France
\hfill\break
\phantom{br.}{\it Home page:}
{\tt http://www.proba.jussieu.fr/pageperso/giacomin/GBpage.html}}
\email{giacomin\@@math.jussieu.fr}
\author{Fabio Lucio Toninelli}
\address{
Laboratoire de Physique, UMR-CNRS 5672, ENS Lyon, 46 All\'ee d'Italie, 
69364 Lyon Cedex 07, France
\hfill\break
\phantom{br.}{\it Home page:}
{\tt http://perso.ens-lyon.fr/fabio-lucio.toninelli}}
\email{fltonine@ens-lyon.fr}
\date{\today}

\begin{abstract}
We present an approach to studying directed polymers in interaction
with a defect line and subject to a force, which pulls them
away from the line. We consider in particular
the case of inhomogeneous interactions.  We first give a formula
relating the free energy of these models to the free energy of the
corresponding ones in which the force is switched off. We then show
how to detect the presence of a re-entrant transition without fully
solving the model.  We discuss some models in detail and show that
inhomogeneous interaction, e.g. disordered interactions, may induce
the re-entrance phenomenon.
\end{abstract}

\maketitle

\section{Introduction}

\subsection{Overview}
Pulling polymers out of potential wells by applying a force $f$ in the
direction orthogonal to the pinning region has been considered at
several instances (see {\sl e.g.} \cite{cf:Iliev_force,
cf:LubNel,cf:Marenduzzo,cf:OTW04} and references therein). The techniques employed
essentially lead to exact (and mathematically rigorous) results for
models with homogeneous interactions (see in particular
\cite{cf:Marenduzzo} and
\cite{cf:OTW04}), while in the inhomogeneous case, notably in the
disordered case, the results are often based on arguments 
  out of mathematical control ({\sl e.g.}, replica and renormalization
  group computations) or on annealed approximations that, in general,
 are not close to the behavior of the quenched system.  
 One of the interesting phenomena that has
been pointed out, at least in some model systems, is the presence of a
re-entrant transition (see in particular \cite{cf:Marenduzzo}, but
this issue is taken up in most of the references we have given). By
this we mean that, for a fixed force $f$, the polymer is pulled out of
the defect line at low temperature and at high temperature, but for
intermediate temperatures it is localized at the defect line, very
much like if the force was not present.  References above are limited
to theoretical work, but all of them are motivated by real
experiments: the literature on real experiments is very rich and here
we single out \cite{cf:Danilo} that deals
with DNA unzipping and contains further references.

Our purpose is to point out that one can directly relate the free
energy in presence of a force to the free energy of the polymer with
free endpoint for a very large class of directed models and that the
presence of a re-entrant transition is easily related to suitable
asymptotic behavior of the expressions appearing in this formula.  As
a consequence, we will give a simple necessary ad sufficient condition
for the re-entrance to take place for very general models (including
disordered models).

In order to be concrete we choose to deal with a precise and rather
limited class of models, but the reader will realize that the method
we present is very general. Extensions are considered explicitly in
Section \ref{sec:generalize}.  Moreover, the phenomenology of the restricted
class of models we consider is already  very rich.

We point out that we focus exclusively on cases in which the polymer
interacts only at the defect line and other relevant cases, like the
case in which bulk disorder is present ({\sl e.g.} \cite{cf:Kafri}),
are not considered here.  On the other hand, it is straightforward to
generalize the content of this note to directed models of copolymers
near a selective interface \cite{cf:RP,cf:GTloc}.

\subsection{A model}
\label{sec:model}
 Let us consider a $(p,q)$-walk, that is a random walk $S:=\{
 S_n\}_{n=0,1, \ldots}$ with independent identically distributed (IID)
 increments $\{S_{n+1}-S_n\}_{n=0,1, \ldots}$, $S_0:=0$ and $S_1$
 taking values in $\{-1, 0,+1\}$, with $\bP (S_1=1)=\bP (S_1=-1)=p/2$
 and $\bP (S_1=0)=q$. We assume $p+q=1$ and $p>0$.  Particularly
 relevant for what follows is the distribution of $\tau:= \inf\{n:\,
 S_n=0\}$: therefore we set $K(n)= \bP (\tau=n)$, along with $\Kbar
 (n) :=
\sum_{j>n} K(j)$.
We point out that the Laplace transform of $K(\cdot)$, that is $\sum_n
\exp(-bn)K(n)$, $b\ge 0$, can be written explicitly and $K(n) $ itself
can be expressed in terms of $\bP (S_k=0)$, $k=1, \ldots, n$, by using
a standard renewal theory formula
$$
 \bP (S_n=0)=\delta_{n,0}+\sum_{j=1}^n K(j)\bP(S_{n-j}=0).
$$
In particular one obtains that $n^{3/2}K(n)$ converges to $c_p:=
\sqrt{p/(2\pi)}$ \cite[App.~A]{cf:RP}, if $p<1$ (an analogous result
holds for the $p=1$ case, see below). Note also that $\Kbar (0)=1$,
since $S$ is recurrent.

Given a sequence $\go:=\{\go_n\}_n$ of real numbers
(the {\sl charges}) we define the new measure $\bP_{N, \go}^{ \gb, f}$
on the space of random walk paths by introducing a Boltzmann
weight ($\gb \ge 0$, $f\ge 0$): 
\begin{equation}
\label{eq:RN}
\frac{\dd \bP_{N, \go }^{ \gb, f}
}{\dd \bP} (S)\, =\, \frac 1{Z_{N, \go}^{ \gb, f}}
\exp \left(
\gb \sum_{n=1}^N \go _n \ind_{S_n=0} \, +\, \gb f S_N
\right).
\end{equation}

We will consider two (wide) classes of charges:
\begin{enumerate}
\item $\go$ is a periodic (deterministic) sequence, {\it i.e.} a sequence
of numbers such that $\go_{n+\tT}=\go_n$ for some positive integer $\tT$
and every $n$. We denote by $\tT(\go)$ the minimal $\tT$ with such a property,
that is the {\sl period}. If $\tT(\go)=1$ then $\go$ is homogeneous.  
\item $\go$ is the realization of a sequence of random variables.
For simplicity we consider only the case of IID variables, therefore the law
of $\go$ is determined by the law of $\go_1$, but our analysis would go through
in the much wider domain of stationary sequences of variables. In the end, 
the properties of the system are determined by the law $\bbP$ of $\go$.
We refer to this case
as {\sl disordered} or {\sl quenched}.
\end{enumerate}

\smallskip 

The existence of the limit
\begin{equation}
\label{eq:fe0}
\lim_{N \to \infty} \frac 1N \log Z_{N, \go}^{ \gb, 0} \, =:\, \tf (\gb),
\end{equation}
is very well known. This follows from elementary super-additivity
arguments in the periodic (and, of course, in the homogeneous) set-up
(see {\sl e.g.}
\cite[Ch.~1]{cf:RP}): in general, the limit depends on $\go$.  In the
disordered case instead one has to be a bit more careful: a precise
statement is for example that if $\bbE \vert \go_1\vert <\infty$ then
the limit in \eqref{eq:fe0} exists $\bbP$-almost surely and also in
the $L^1(\bbP)$ sense. Moreover the right-hand side is in principle a
random variable, but it turns out to be degenerate, that is almost
surely independent of $\go$.  This property, usually referred to as
{\sl self-averaging}, is well understood in this context (see
\cite[Ch.~4]{cf:RP} for proofs and overview of the literature).  We
remark that
\begin{equation}
Z_{N,\go}^{ \gb, 0}\ge \bE\left[
S_n\neq 0, \text{ for } n=1,2, \ldots, N \right] \, =\, \Kbar (N) \, \ge \, 
c N^{-1/2},
\end{equation}
for some $c>0$, so that $\tf(\gb)\ge 0$ for every $\gb$.
Much literature  has been spent on this model and about the fact
that $\tf(\gb)>0$ corresponds to localized regime, that is to the case
in which the typical trajectories are tightly bound to the defect
line (cf., e.g., \cite{cf:GTloc} and
\cite[Ch. 7]{cf:RP}).  On the other hand $\tf (\gb)=0$ corresponds to a delocalized
regime. One should however distinguish between the critical point
$\beta=\beta_c =\sup\{\beta:\tf(\beta)=0\}$ and the truly delocalized
region where $\beta<\beta_c$.  Several results are available also on
the delocalized regime (see \cite{cf:RP} and references therein, in
particular \cite{cf:GTdeloc} for the disordered case) and they say,
very roughly, that very few visits are paid by the polymer to the
defect line.

A relevant difference between the two classes of charges we consider
is that in the first case $\tf(\gb)$ is explicitly known. To be more precise
in the periodic case
$\tf(\gb)$ can be  expressed in terms of the leading eigenvalue of 
a suitable $\tT(\go)\times \tT (\go)$ Perron--Frobenius matrix \cite{cf:BG,cf:CGZ3} (see also Appendix \ref{sec:AppT}):
computing such an eigenvalue becomes harder and harder for
larger periods, but the problem trivializes in the homogeneous case. 
In the disordered case instead only estimates on $\tf(\gb)$ are known.

We point out also that for every $y\in \R$ the limit
\begin{equation}
\lim_{N\to \infty} \frac 1N \log \bE\left[ \exp \left(yS_N\right);
\, S_n \neq 0 \text{ for } n=1,2, \ldots, N\right] \, :=\, \tg(y),
\end{equation}
exists. Actually, the explicit value of $\tg(\cdot)$ is easily computed
by applying the reflection principle \cite[App. A]{cf:RP} and one obtains
$\tg(y)=\log(p \cosh(y) +q)$. 

\smallskip

For the model with the force we have:

\bigskip

\begin{proposition}
\label{th:decomp}
For every $\gb$ and $f$ the limit, that we denote by $\tf(\gb,f)$, of
the sequence $\left\{ (1/N) \log Z_{N, \go}^{ \gb, f}\right\}_N$
exists ($\, \bbP(\dd \go) $-almost surely (a.s.), in the disordered
case). Moreover we have the formula
\begin{equation}
\label{eq:decomp}
\tf(\gb,f) \, =\, \max \left(\tf (\gb) , \tg(\gb f)\right).
 \end{equation}
\end{proposition}

\bigskip

A line of non-analytic points of $\tf(\cdot, \cdot)$ is therefore
evident: for every $\gb$ we set $f_c(\gb):=\gb^{-1} \tg^{-1} \left(
\tf(\gb)\right)\in [0, \infty)$ (we are looking at  $\tg(\cdot)$ 
as a function, and a bijection, from $[0, \infty)$ to $[0, \infty)$). Therefore for every $\gb$
\begin{equation}
\tf(\gb, f) \, =\, \begin{cases}
 \tf(\gb) &\text{if } f\le f_c(\gb),\\
 \tg(\gb f) &\text{if } f\ge f_c(\gb).
\end{cases}
\end{equation}

It is quite easy to get convinced that this non-analyticity
corresponds to a localization--delocalization transition: the most
interesting case is when $\tf(\gb)>0$, otherwise the system is already
delocalized at $f=0$. By convexity, $\partial _f \tf (\gb,
f)>0$ if $f>f_c(\beta)$ (and if $\partial _f \tf (\gb, f)$ exists).
This directly implies that $\lim_{N \to \infty} \bE_{N, \go}^{\gb,f
}S_N/N >0$ (and the same statement holds even if $\partial _f \tf
(\gb, f)$ does not exist, but one has to replace the $\lim$ with
$\liminf$).  For $f<f_c(\gb )$ instead $\lim_{N \to \infty} \bE_{N,
  \go}^{\gb,f }S_N/N =0$.  These are distinctive marks respectively of
localization and delocalization (for sharper results we refer
to \cite{cf:Nthese}).

It is also worthwhile to observe that formula \eqref{eq:decomp}
yields that $\partial_f \tf (\gb, f)$ is discontinuous at $f=f_c(\gb)$
for $\gb>0$. So the transition is of first order (as argued in much of the
previous literature). The underlying mechanism has also been expoited in \cite{cf:A2}.

\smallskip

We now turn to the issue of the existence of a re-entrant transition.
As mentioned above, this refers to the fact that for some fixed $f$
the system undergoes (at least) two phase transitions as the
temperature $T=1/\beta$ is increased from zero to infinity. More
precisely, two cases are observed: one can either observe a pattern of the
type localized-delocalized-localized by increasing $T$, or the
opposite one: delocalized-localized-delocalized. 

Of course, re-entrance is equivalent to the non-monotonicity of $f_c$
as a function of $\beta$ and a sufficient condition for it is that the
difference $f_c(\infty)-f_c(0)$ has the opposite sign as
$\partial_\beta f_c(0)$.  As we show in the following, these
quantities are related to the asymptotic behavior of $\tf(\beta)$ for
$\beta\to0$ and $\beta\to\infty$, which in many cases (including
quenched disordered situations) can be easily computed without fully
solving the model.

We wish to emphasize that, while in principle the occurrence of 
\begin{eqnarray}
\label{eq:suffcond}
\partial_\beta f_c(0)[f_c(\infty)-f_c(0)]<0, 
\end{eqnarray} 
is just a sufficient condition for re-entrance, it turns out
numerically in the cases we have checked ({\sl cf.} 
Fig.~\ref{fig:homog} and \ref{fig:T}) that when \eqref{eq:suffcond} is not verified
the critical force is monotone in $\beta$ and re-entrance is absent.

\section{The homogeneous case}
\label{sec:hom}

This section has the two aims:

\begin{enumerate} 
\item
generalizing the results
of  the previous literature. For example in \cite{cf:OTW04} only the cases $q=1/3$
and $q=0$ are considered and no re-entrance is observed, while for different values of
$q$ re-entrance does appear;
\item
  providing the general scheme that we follow
also in the next sections,  even if the models of this section
are exactly solvable (but we will not solve them).
\end{enumerate}
\smallskip

First of all note that, if $\go_n\equiv -c<0$ the polymer is
delocalized already in absence of the force, and $f_c(\beta)=0$ for
every $\beta\ge0$, an uninteresting situation.
Therefore, in the homogeneous case, 
we will assume 
that $\go_n\equiv c>0$ and,  without loss of generality, $c=1$.

\smallskip
\begin{figure}[!h]
\begin{center}
\leavevmode
\epsfysize = 9 cm
\psfragscanon
\psfrag{fc}[c][l]{$f_c(\gb) $}
\psfrag{betainv}[c][l]{$1/\gb $}
\psfrag{2.0}[c][c]{\tiny $2.0$}
\psfrag{1.0}[c][c]{\tiny $1.5$}
\psfrag{1.0}[c][c]{\tiny $1.0$}
\psfrag{0.5}[c][c]{\tiny $0.5$}
\psfrag{0.0}[c][c]{\tiny $0.0$}
\psfrag{0.2}[c][c]{\tiny $0.2$}
\psfrag{0.4}[c][c]{\tiny $0.4$}
\psfrag{0.6}[c][c]{\tiny $0.6$}
\psfrag{0.8}[c][c]{\tiny $0.8$}
\psfrag{1.2}[c][c]{\tiny $1.2$}
\psfrag{1.4}[c][c]{\tiny $1.4$}
\psfrag{0}[c][c]{\tiny $0$}
\psfrag{1}[c][c]{\tiny $1$}
\psfrag{2}[c][c]{\tiny $2$}
\psfrag{3}[c][c]{\tiny $3$}
\psfrag{4}[c][c]{\tiny $4$}
\psfrag{6}[c][c]{\tiny $6$}
\psfrag{8}[c][c]{\tiny $8$}
\psfrag{10}[c][c]{\tiny $10$}
\psfrag{a}[c][c]{{\bf (a)}}
\psfrag{b}[c][c]{{\bf (b)}}
\epsfbox{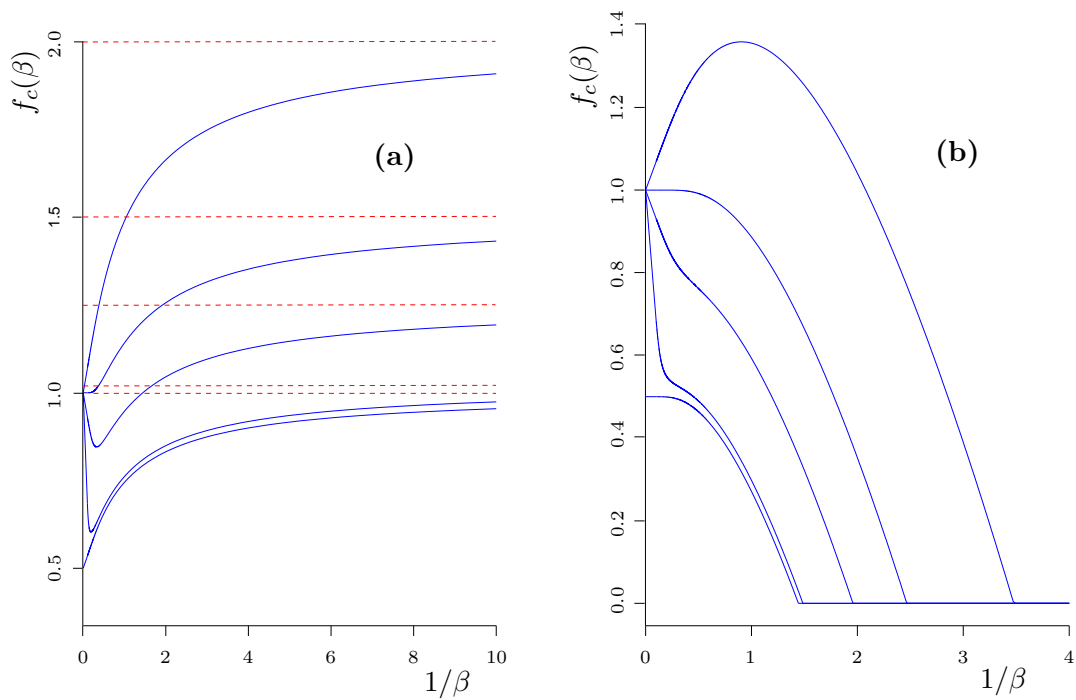}
\end{center}
\caption{\label{fig:homog} Critical force $f_c$ as function of 
  $1/\beta$, for the homogeneous model. {\bf (a)} Without hard wall. The
  curves correspond $q=1/2$, $1/3$, $1/5$, $1/50$ and $0$, from top to bottom. Re-entrance is
  present if and only if $0<q<1/3$. The critical force $f_c (\gb)$ is computed  by using 
  \eqref{eq:decomp}, where $\tf(\beta)$ is the  solution
  of $\sum_n K(n)\exp(-\tf(\beta)n)=\exp(-\beta)$ \cite[Proposition
  1.1]{cf:RP}.  {\bf (b)} With hard wall (the values of $q$ are the
  same as in (a)).
  Re-entrance is present if and only if $q>1/3$. In this case, 
$\tf(\beta)$ is given by the solution of 
$\sum_n K^+(n)\exp(-\tf(\beta)n)=\exp(-\beta)$, where $K^+(n)=K(n)/2$ 
if $n\ge 2$ and $K^+ (1)=K(1)$.
}
\end{figure}

 The behaviors of the critical force for
$\beta\searrow 0$ and for $\beta\to\infty$ are easily obtained:
\begin{proposition}
\label{th:hNR}
For the homogeneous model, {\it i.e.}
\eqref{eq:RN} and $\go_n \equiv 1$,  with $p<1$ one has
\begin{eqnarray}
\label{eq:bt0}
\lim_{\beta\searrow0}
f_c(\beta)\, =\,  \frac1p ,
\end{eqnarray}
and
\begin{eqnarray}
\label{eq:pqRW}
  f_c(\beta)\stackrel{\beta\to\infty}= 
 1+\frac 1\gb \left( \log q -\log(p/2) \right) +o\left(\frac 1\gb \right).
\end{eqnarray}
In particular, one has re-entrance if and only if   $2/3<p<1$.
As for $p=1$, one has $f_c(\beta)\to1$ for $\beta\to0$ and 
\begin{equation}
\label{eq:SRW}
f_c (\gb)\, \stackrel{\beta\to\infty}=\, \frac 12 + \frac 1{2\gb}\log2 
+o\left(\frac 1\gb \right).
\end{equation}
Re-entrant behavior is not observed  in this case.
\end{proposition}
See also Figure \ref{fig:homog}(a), were we plot $ f_c(\gb)$ as a
function of $1/\beta$, for different choices of $q$.  Note that the
transition $p\to1$ to the simple random walk case is singular.

\smallskip

\noindent
{\it Proof of Proposition \ref{th:hNR}}.  Consider first the case
$p<1$. Since $K(n)\sim c_p n^{-3/2}$, $c_p=\sqrt{p/(2\pi)}$,  for $\beta\searrow 0$
one has \cite[Theorem 2.1]{cf:RP}
\begin{eqnarray}
  \label{eq:fbto0}
\tf(\beta)\sim \frac{\beta^2}{(2c_p\Gamma(1/2))^2}.  
\end{eqnarray}
 This, together with a Taylor expansion of $\tg(y)$
around $y=0$, immediately gives \eqref{eq:bt0}.

 The large $\beta$ behavior can also be easily captured without
explicit computations: for $p<1$, $\tf (\gb) = \gb +\log K(1)+o(1)$ as
$\gb \to \infty$, and $K(1)=q=1-p$.  This corresponds simply to the
fact that for $\gb$ large the dominant trajectories are the ones such
that $\vert \{ n:\, S_n \neq 0\}\vert\ll N$.  Since $\tg (y)= y+ \log
(p/2)+o(1) $ as $y \to \infty$, one finds  \eqref{eq:pqRW}.

For $p=1$, \eqref{eq:fbto0} is modified in this case into
$\tf(\beta)\sim \beta^2/2$ (this follows from \cite[Theorem
2.1]{cf:RP} plus the fact that, for the simple random walk, $K(2n)\sim
n^{-3/2}\sqrt{1/(4\pi)}$ \cite[Ch. III]{cf:Feller1} and
$K(2n+1)=0$). As a consequence, $f_c(\beta)\to 1$ for $\beta\to0$.  As
for the $\beta\to\infty$ behavior notice that, $\tf (\gb) =
\gb/2 +(\log K(2))/2+o(1)$ and $K(2)=1/2$, so that  \eqref{eq:SRW}
follows. 

The statements about re-entrance easily follow from 
\eqref{eq:bt0}-\eqref{eq:SRW}.
\qed

\subsection{The model with hard wall repulsion}
We conclude this section by showing that the phase diagram and  the re-entrance phenomenon are
strongly model dependent.  To this purpose, we modify the model by
inserting a hard wall condition (like in \cite{cf:OTW04}), which corresponds to
inserting in the right-hand side of \eqref{eq:RN} the indicator
function of the event $\{S:S_n\ge0$ for $n=1,\ldots,N\}$.  Of course,
the partition function and the free energy will be in general
modified. We consider for conciseness only the case $p<1$.  
\begin{proposition}
\label{th:propwall}
  For the homogeneous model with hard-wall repulsion and $p<1$, 
there exists $\beta_0>0$ such that $f_c(\beta)=0$ for $\beta<\beta_0$. 
Moreover, the large $\beta$ behavior of $f_c(\beta)$ is still given
by  \eqref{eq:pqRW}. Re-entrance takes place for $p<2/3$.
\end{proposition}

\smallskip

\noindent
{\it Proof of Proposition \ref{th:propwall}}.  In presence of the hard
wall, for $\beta$ sufficiently small but finite one has $\tf(\beta)=0$
(cf.  for instance \cite[Section 1.2]{cf:RP}), so that $f_c(\beta)=0$.
This is rather intuitive: even in absence of the pulling force, the
entropic repulsion effect provided by the wall is enough to delocalize
the polymer. As for the $\beta\to\infty$ limit, since the dominant
trajectory $S_n\equiv0$ is allowed by the hard wall constraint, one
finds again $\tf(\beta)= \beta+\log K(1)+o(1)$, and Equation
\eqref{eq:pqRW}. Occurrence of re-entrance for $p<2/3$ immediately follows
from the small- and large-$\beta$ behavior of the critical force.
\qed

\smallskip

Observe that for the model with hard wall repulsion the situation
is somewhat reversed with respect to the previous case: one has
re-entrance for $p<2/3$ and no re-entrance for $2/3<p<1$ 
(see also Fig.~\ref{fig:homog}(b)).

\section{The periodic case}
\label{sec:T}

\begin{figure}[!h]
\begin{center}
\leavevmode
\epsfysize = 8 cm
\psfragscanon
\psfrag{fc}[c][l]{$f_c(\gb) $}
\psfrag{T}[c][l]{$T (=1/\gb)$}
\psfrag{0.8}[c][c]{$f_c(\gb)$}
\psfrag{0.0}[c][c]{\tiny $0.0$}
\psfrag{0.2}[c][c]{\tiny $0.2$}
\psfrag{0.4}[c][c]{\tiny $0.4$}
\psfrag{0.6}[c][c]{\tiny $0.6$}
\psfrag{0}[c][c]{\tiny $0$}
\psfrag{2}[c][c]{\tiny $2$}
\psfrag{4}[c][c]{\tiny $4$}
\psfrag{6}[c][c]{\tiny $6$}
\psfrag{8}[c][c]{\tiny $8$}
\psfrag{10}[c][c]{\tiny $10$}
\psfrag{seq30}[c][c]{\tiny $1 0 0 0 1 0 0 1 1 1 1 1 0 1 0 1 1 1 1 1 1 0 1 1 1 0 0 1 1 0 0 0 1 0 1 0 1 0 0 1 1 1 1 1 0 1 1 0 0 1$}
\psfrag{seq25}[c][c]{\tiny $0 1 0 1 0 0 1 1 1 0 0 0 0 1 1 0 0 1 1 0 1 1 1 1 0 1 0 1 1 1 1 0 1 0 0 0 0 0 1 1 0 1 0 0 1 1 0 0 0 1$}
\psfrag{seq22}[c][c]{\tiny $0 1 0 1 0 1 1 0 0 0 0 0 0 0 0 1 0 1 0 1 1 1 1 0 0 1 0 1 1 0 1 0 1 0 0 1 0 0 0 1 1 1 0 1 0 0 1 0 1 0$}
\epsfbox{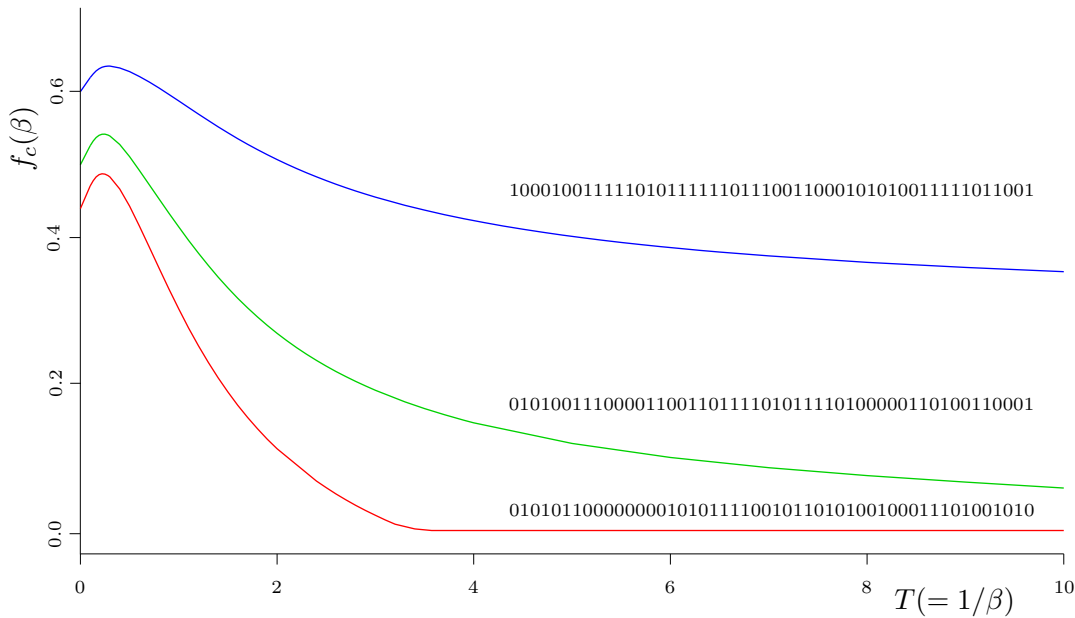}
\end{center}
\caption{\label{fig:T} Critical force $f_c$ as function of 
  $1/\beta$, in the case of periodic charge distribution 
  and $q=1/3$. In the three cases $\tT (\go)= 50$
and the charges have been chosen at random, {\it i.e.} each
sequence $\{\go_n \}_{n=1, \ldots, 50}$, $\go_n\in\{-1,1\}$,
is equiprobable
(the explicit sequences 
$\{(\go_n+1)/2\}_{n=1, \ldots, 50}$ are given next to the corresponding curves):
for the top curve $\vert \{ n :\, \go_n =+1\}\vert =30$,
while for the bottom one $\vert \{ n :\, \go_n =+1\}\vert =22$.
The intermediate curve corresponds instead to 
the case in which there are as many positive charges
as negative, and in fact the curve approaches $0$ 
for large temperatures. As explained in the text, these 
graphs give also strong hints on the behavior
of disordered systems. }
\end{figure}

We start by recalling that the  solution of 
the periodic case can be reduced to a finite dimensional, 
in fact $\tT(\go)$--dimensional,
problem (\cite{cf:BG,cf:CGZ3,cf:RP}): 
by solving numerically this finite dimensional problem
we have drawn the curves in Fig.~\ref{fig:T}.
However, from the $\tT(\go)$--dimensional problem one can extract
many analytic features too. Here we concentrate on
small and large $\gb$ behavior, since they suffice to
highlight part of the variety of observed phenomena:
we collect and discuss here the results, see App.~\ref{sec:AppT} for
 the proofs. 
 \smallskip 
 
 Just a bit of notation for the following result (for sake of conciseness,
  we restrict
 our attention to the case without hard-wall repulsion).  Set
 $\ell_0:=\min \{n:\, \go_n=+1\}$ (we are implicitly assuming that
 $\go_n $ is not equal to $-1$ for all $n$) and $\ell_{i+1}:=
 \inf\{n>\ell_i:\, \go_n =+1\}$. Set also $\iota (\go):= \min\{i:\,
 \ell _i > \tT(\go)\}$.

\medskip

\begin{proposition}
\label{th:Tresults}
In the periodic case we have the small temperature
behavior 
\begin{equation}
\label{eq:Tres1}
f_c (\gb)\,\stackrel{\gb \to \infty}=\,
\frac 1{\tT (\go)}\sum_{n=1}^{\tT(\go)} (2\go_n -1)
+ \frac 1 \gb  \left(
\frac 1{\tT (\go)} \sum_{i=1}^{\iota (\go)} \log K(\ell_i -\ell_{i-1}) - \log(p/2)
\right) + o\left(1/ \gb \right), 
\end{equation}
and at high temperatures ($\gb \searrow 0$)
\begin{equation}
\label{eq:Tres2}
f_c (\gb)  \, =\, 
\begin{cases}
  \frac 1{p \tT (\go)} \sum_{n=1}^{\tT(\go)} \go_n \, +
O(\gb)
& \text{if } \sum_{n=1}^{\tT(\go)} \go_n > 0,
\\
\frac 1{2p} \gb
 +o(\gb) & \text{if } \sum_{n=1}^{\tT(\go)} \go_n = 0,
 \\
 0 & \text{if } \sum_{n=1}^{\tT(\go)} \go_n< 0 \text{ and } \gb \le \gb_0,
 \end{cases}
\end{equation}
where $\gb_0 := \sup\{\gb:\, \tf (\gb)=0\}$, and $\gb_0 \in (0, \infty)$
if $\sum_{n=1}^{\tT(\go)} \go_n< 0$.
\end{proposition}
\medskip
 
 Of course such a result is sufficient in order to check our sufficient
 condition \eqref{eq:suffcond} for the occurrence of a re-entrant
 transition, for a given periodic charge sequence $\omega$. On the
 other hand, a full characterization of the critical curve in general
 requires some numerical computations (but we stress once again that
 they are just finite dimensional computations).  Note however, by
 comparing with the homogeneous case, that the inhomogeneous character
 of the charges induces a re-entrant transition for example if $q=1/3$
 (but of course it can induce it also for $q <1/3$).
 
 An important feature resulting from  Prop.~\ref{th:Tresults}
 (but also from Fig.~\ref{fig:T})
 is the strong dependence on $\go$. In particular the sign of the mean
 over a period leads to drastically different behaviors for large temperatures. 
 It should be however pointed out that if $\go_1, \ldots, \go_{\tT}$ are
 sampled in a IID fashion, like in the quenched disordered case, then in the limit of $\tT \to \infty$ 
 the free energy of the periodic model converges
 $\bbP (\dd \go)$--a.s. toward the free energy
 of the disordered model  \cite[Th.~4.5]{cf:RP}. One then sees directly that
 such a result implies the convergence of the critical force
 of the model with period $\tT$ to the critical force of the corresponding 
 disordered model.
 In  Fig.~\ref{fig:T}
we have plotted a case in which $\sum_{n=1}^{\tT} \go_n=0$
and two cases deviating above and below the mean by more than
one standard deviation. 


\section{The disordered case}
Let us now have a look at the disordered model.
For the sake of conciseness, we consider only the case $p<1$ and
$\bbP( \go_1=+1)= \bbP( \go_1=-1)=1/2$. 

It is known that in this case the model with $f=0$ and without hard
wall repulsion is localized for every $\gb>0$ (see
\cite[Ch.~5]{cf:RP}).  The  behavior of the free energy  is not
fully under  control, but one can prove the following:
 \begin{proposition}
\label{th:disord}
For the disordered model, $f_c(\beta)\to0$ for $\beta\to0$. 
As for large $\beta$, one has
\begin{equation}
\label{eq:bounds2}
 f_c(\gb) \, \stackrel{\beta\to\infty}=\,\frac 12 + \frac 1\gb \left(
\frac 12 \sum_{j=1}^\infty 2^{-j} \log K(j) -\log(p/2)
\right)+ o\left( \frac 1\gb\right)
\end{equation}
so that re-entrance is observed as soon as 
\begin{equation}
\frac 12 \sum_{j=1}^\infty 2^{-j} \log K(j) \, >\, \log(p/2),
\end{equation}
that is for $q>0.1994...$.
 \end{proposition}
 
\smallskip

{\sl Proof of Proposition \ref{th:disord}}.
For $\beta\to0$, the annealed bound
\begin{equation}
\tf(\beta)\le \lim_{N\to\infty} \frac1N \log \bbE Z_{N,\go}^{\beta,0},
\end{equation}
 is already sufficient to prove that $f_c(\beta)\to0$.  In fact,  the
right-hand side is just the free energy of the homogeneous model where
$\beta$ is replaced by $\log(\cosh\beta)$, and therefore (recall
\eqref{eq:fbto0} and discussion thereafter) it behaves like
$\beta^4/(8p)$ for $\beta$ small.  This immediately implies that
$f_c(\beta)\to 0$ for $\beta\to0$.

\smallskip

\begin{rem}
In  \cite{cf:A} it is actually proven that $\tf (\gb) \stackrel{\gb \searrow 0} {\sim} \gb ^4/(8p)$,
so that $f_c(\gb) \stackrel{\gb \searrow 0} {\sim} \gb/(2 \sqrt{p})$.
\end{rem}

\smallskip

As for large $\beta$, we need the following: 

\smallskip

\begin{lemma}
\label{th:19}  
For every $\gep>0$ there
 exists $\gb_\gep$ such that for $\gb\ge \gb_\gep$ we have
\begin{equation}
\label{eq:bounds1}
\left\vert
\tf (\gb) -\frac \gb 2 -
\frac 12 \sum_{j=1}^\infty 2^{-j} \log K(j) \right\vert
\, \le\, \gep.
\end{equation}
\end{lemma}
\smallskip

Note that \eqref{eq:bounds2} is a direct consequence of 
Lemma~\ref{th:19} and the proof of the latter can
be found 
 in Appendix \ref{sec:A1}.  Here, we
just notice that  \eqref{eq:bounds1} shows how the situation is
different from the homogeneous case, where one has rather
$\tf(\beta)=\beta+\log K(1)+o(1)$. In fact also in the disordered 
situation  the dominant
configurations are those for which $S_n=0$ for every $n$ such that
$\omega_n=+1$ and $S_n\ne0$ otherwise. However, now 
the average density of $+1$ charges is $1/2$ (whence the leading
term $\beta/2$) and the distance
between two successive positive charges
 is random and can take any value $j=1,2, \ldots$ with geometric
probability $2^{-j}$.
\qed

\smallskip

As explained in Section~\ref{sec:T}, we expect the graph
of $f_c(\cdot)$ to be close to the
intermediate curve in Fig.~\ref{fig:T}.

\subsection{The model with repulsion}
\label{sec:disrep}
Like in the homogeneous case, if we add the hard-wall repulsion then
for $\beta$ sufficiently small $\tf(\beta)=f_c(\beta)=0$ 
(see for instance \cite[Proposition 5.1]{cf:RP}).
Moreover, a look at the proof of \eqref{eq:bounds1} shows that
Lemma~\ref{th:19} still holds in this case, provided that
$K(n)$ is replaced by $K^+(n)$ defined as $K^+(1):=K(1)$ and
$K^+(n):=K(n)/2$ for $n\ge2$.  This is just related to the fact that,
if $n\ge2$, of all the possible trajectories $\{S_0,\ldots,S_n\}$ of
the $(p,q)$ walk satisfying $S_0=S_n=0, S_i\ne0$ for $1\le i<n$, only
half survive the introduction of the hard wall constraint.  Therefore, 
\eqref{eq:bounds2} still holds with $K(.)$ replaced by $K^+(.)$
and a sufficient condition for having a re-entrant transition is 
\begin{equation}
\frac 12 \sum_{j=1}^\infty 2^{-j} \log K^+(j) \, >\, \log(p/2).
\end{equation}
Numerically, this corresponds to $q>0.2838\ldots$.

\section{Generalizations}
\label{sec:generalize}

The procedure presented in this paper to study the phase diagram
and the occurrence of a
re-entrant phase transition applies well beyond the case of the
$(p,q)$-walks.  We have seen in fact that all what one needs to know is
the asymptotic behavior of $\tg(y)$ for small and large $y$, and the
asymptotic behavior of $\tf(\beta)$ for small and large $\beta$.

Just to show an example of generalization, consider the case of a
directed polymer in $(2+1)$ dimensions with homogeneous ($\go_n\equiv
1$) pinning attraction to a defect line.  
In particular, we will assume that $S_n\equiv
(S_n^{(1)},S_n^{(2)})
\in \Z^2$ and $\{S_n\}_{n=0,1,\ldots}$
is the two-dimensional simple random walk where $S_{n+1}$ is chosen
uniformly among the 4 neighbors of $S_n$. In this section, it will be
understood that $K(n)$ is the probability that the first return to
zero of the two-dimensional random walk occurs at time $n$ (in the
previous sections the same symbol was used for the analogous quantity
referring to the $(p,q)$-random walk).

Again, the Boltzmann weight is defined as in  \eqref{eq:RN},
with the only difference that the term $f S_N$ is replaced by, say, $f
S_N^{(1)}$. This corresponds to assuming that the force is pulling in
the direction ``$1$'' (orthogonal to the defect line).  Then,
Proposition \ref{th:decomp} still holds, provided that in the
definition of $\tg(y)$ one puts $y S_N^{(1)}$ instead of $y S_N$.

The analogue of Proposition \ref{th:hNR} is the following:
\begin{proposition}
  \label{th:2+1}
For the homogeneous $(2+1)$-dimensional model one has $ f_c(\beta)\to0$ for
$\beta\to0$ and
\begin{eqnarray}
\label{eq:does}
  f_c(\beta)\stackrel{\beta\to \infty}=\frac12+\frac{1}\beta \log 2
+o\left(\frac1\beta\right).
\end{eqnarray}
Re-entrance does take place.
\end{proposition}

\smallskip

\noindent
{\it Proof of Proposition \ref{th:2+1}}. It is not difficult to
prove that for the two-dimensional simple random walk
\begin{eqnarray}
\label{eq:g2+1}
\tg(y)=\log(\cosh(y)+1)-\log 2.
\end{eqnarray}
Indeed, first of all one has
\begin{eqnarray}
\label{eq:glb}
 \bE[\exp(y S_N^{(1)});
S_n\ne 0 \mbox{\;\;for\;\;}n=1,2,\ldots,N]&\le& 
 \bE[\exp(y S_N^{(1)})]= \bE[\exp(y S_1^{(1)})]^N
\\\nonumber
&=&
(1/2+(1/2)\cosh y)^N
\end{eqnarray}
for every $N$, so that 
\begin{eqnarray}
\label{eq:ubg}
\tg(y)\le \log (\cosh(y)+1)-\log 2. 
\end{eqnarray}
To get the complementary  lower bound, note that 
\begin{eqnarray}
\label{eq:gub}
 \bE[\exp(y S_N^{(1)});
S_n\ne 0 \mbox{\;\;for\;\;}n=1,\ldots,N]
\ge \bE[\exp(y S_N^{(1)});
S_n^{(1)}\ne 0 \mbox{\;\;for\;\;}n=1,\ldots,N].
\end{eqnarray}
 The quantity one is averaging in the right-hand side of
 \eqref{eq:gub} involves only $\{S^{(1)}_n\}_{n\ge0}$, which is just a
 $(p,q)$-walk with $p=1/2$. Therefore, from the knowledge of $\tg(y)$
 for the $(p,q)$-walk (cf. Section \ref{sec:model}) one obtains
 $\tg(y)\ge \log(1/2+(1/2)\cosh(y))$ which concludes the proof of
 \eqref{eq:g2+1}.

Next, it is known \cite{cf:2d} that $K(2n)\sim
c/(n(\log n)^2)$ with $c>0$ while $K(2n+1)=0$. As a consequence
(cf. \cite[Theorem 2.1]{cf:RP}), $\tf(\beta)$ vanishes for $\beta\to0$
faster than any power of $\beta$. This, together with 
the expansion $\tg(y)=y^2/4+o(y^2)$ for
$y\to0$, implies that $f_c(\beta)\to0$. 
As for the large-$\beta$ behavior, in analogy with the discussion in 
Section \ref{sec:hom}, a look at the dominant trajectories gives $\tf(\beta)=
\beta/2+(\log K(2))/2+o(1/\beta)$, where now $K(2)=1/4$.
Since $\tg(y)=y-2\log2+o(1)$ for $y\to\infty$,  
\eqref{eq:does} follows from Proposition \ref{th:decomp}.
\qed
\medskip

The disordered $(2+1)$-dimensional model where $\go_n$ are IID symmetric
random variables $\omega_n=\pm1$ 
can also be analyzed in our framework, with the following
result:
\begin{proposition}
\label{th:21disord}
For the disordered $(2+1)$-dimensional pinning model, $f_c(\beta)\to0$ for
$\beta\to0$ and
\begin{eqnarray}
\label{eq:gfp}
f_c(\beta)\stackrel{\beta\to\infty}=
\frac14+\frac1\beta\left[2\log2+\frac14\sum_{j\ge1}
2^{-2j}\log K(2j)\right]+o\left(\frac1\beta\right).
\end{eqnarray}
Re-entrance is  observed. 
\end{proposition}
The proof of Proposition \eqref{th:21disord} is essentially identical
to that of Proposition \ref{th:disord}. The reason why the factors
$1/2$ of \eqref{eq:bounds2} are replaced by $1/4$ is just that our
two-dimensional walk can touch the defect line only for $n$ even, and
therefore for large $\beta$ dominant trajectories will touch half of
the total positive charges, i.e., approximately $N/4$ of them.  The
occurrence of re-entrance follows from a numerical evaluation
of the constant multiplying $1/\beta$ in \eqref{eq:gfp}, which
turns out to be equal to $1.2427\ldots$.

\appendix
\section{Proofs and technical estimates}

\subsection{Proof of Proposition~\ref{th:decomp} and Lemma~\ref{th:19}}
\label{sec:A1}
{\it Proof of Proposition~\ref{th:decomp}.}
Decompose the trajectory according to the
location of the last visit of $S$ to the origin before $N$
(including $N$) getting thus by the Markov property of $S$
\begin{multline}
Z_{N, \go}^{ \gb, f}\, = \\
\sum_{m=0}^N 
\bE\left[ \exp \left(
\gb \sum_{n=1}^m \go _n \ind_{S_n=0} 
\right); \, S_m =0\right]
\bE\left[
\exp \left( \gb f S_{N-m}
\right); \, S_k\neq 0, k=1, \ldots, N-m \right].
\end{multline}
The right-hand side has the form
$\sum_{m=0}^N a_m(\go) b_{N-m}$,
with $a_0(\go)=b_0=1$, $a_m(\go)= \exp( (\tf(\gb)+o(1)) m)$
for $m\to \infty$, $\bbP(\dd\go)$-a.s. (the presence 
of the indicator function of the event
$S_m=0$ is irrelevant, see {\it e.g.} \cite[Remark 1.2]{cf:RP}),
and $b_m = \exp((\tg (\gb f)+o(1)) m)$, again for $m\to \infty$.
At this point we remark that 
$Z_{N, \go}^{ \gb, f} \ge \max (a_N(\go), b_N)$
and therefore 
$\liminf_{N \to \infty} (1/N) \log Z_{N, \go}^{ \gb, f} $ is bounded
below by $ \max \left(\tf (\gb) , \tg(\gb f)\right)$.
For the opposite inequality
we notice that for every $\gep>0$ there exists
$A(\go)$ such that 
$a_n(\go) \le A(\go) \exp \left( ( \tf (\gb) +\gep) n \right)$
for every $n$. Analogously, $b_n(\go) \le B \exp \left( ( \tg (\gb) +\gep) n \right)$
for some constant $B$ and every $n$.
Therefore
\begin{equation}
Z_{N, \go}^{ \gb, f}\, \le  A(\go) B
\sum_{m=0}^N  \exp \left(  \tf (\gb) m +\tg(\gb f) (N-m) +\gep N \right),
\end{equation}
so that 
\begin{equation} 
Z_{N, \go}^{ \gb, f}\, \le \,
(N+1)
A(\go) B  \exp\left( \max \left(\tf (\gb) , \tg(\gb f)\right) N+ \gep N\right), 
\end{equation}
for every $N$ and, since $\gep>0$ can be chosen
arbitrarily small,  the proof is complete. \qed

\bigskip

\noindent
{\it Proof of Lemma~\ref{th:19}.}
We set
\begin{equation}
Q\, :=\,\frac 12 \sum_{j=1}^\infty 2^{-j} \log K(j),  
\end{equation}
and we  separate the proof in lower and upper bound. 

 For the lower bound we select  
the $S$ trajectories hitting $0$ if and only if the charge is $+1$
on that site. This yields
\begin{equation}
\label{eq:LB}
Z_{N, \go}^{ \gb, 0}\, \ge \, \left(\prod_{j=1}^{\cN_N(\go)} \exp(\gb) K(\ell_j) \right)
\Kbar \left(N- \sum_{j=1}^{\cN_N(\go)}\ell_j \right), 
\end{equation}
where $\ell_1:= \inf\{n>0: \, \go_n=1\}$, $\ell_{k+1}:= \inf\{n>0: \, \go_{n+\sum_{j=1}^k \ell_j}=1\}$
and $\cN_N(\go):= \max\{k:\, \sum_{j=1}^k \ell_j \le N\}$.
Note that $\{\ell_j\}_j$ is an IID sequence of geometric random variables
of parameter $1/2$.
 By taking the logarithm and dividing by $N$ both
 sides in \eqref{eq:LB}, in the limit as $N \to \infty$ we get
 \begin{equation}
 \label{eq:withlimsup}
 \tf(\gb) \, \ge \, \gb\limsup_{N \to \infty} \frac 1N\cN_N (\go)
 +\limsup_{N \to \infty} \frac 1N \sum_{j=1}^{\cN_N(\go)}\log K(\ell_j),
 \end{equation}
where we have used the fact that $\Kbar \left(N- \sum_{j=1}^{\cN_N(\go)}\ell_j \right)
\ge \Kbar (N)$ and $(\log \Kbar (N))/N \to 0$ as $N \to \infty$.
By the classical Renewal Theorem $\cN_N(\go)/N$
actually converges $\bbP (\dd \go)$-a.s. to 
$1/\bbE[\ell_1] = 1/2$ and, in turn, by the
Strong Law of Large Numbers also the second
superior limit in the right-hand side of \eqref{eq:withlimsup}
is an almost sure limit and it is equal to $\bbE[ \log K(\ell_1)]/\bbE[\ell_1]$,
which coincides with $Q$.
This concludes the proof of the lower bound.
\medskip

For the upper bound we define $A_{N , \go}:=\{n:\, \go _n=+1\}
\cap [1,N]$ and 
for every $\gep>0$ and any
realization of $\go$ we define the set of trajectories
\begin{equation}
\gO_{N, \go, \gep }\, :=\, \left\{S:\, \left\vert \{1\le n\le N :\, S_n=0\} \bigtriangleup
A_{N , \go} \right \vert \, \le\, \gep \vert A_{N , \go}\vert
\right\},
\end{equation}
where $\bigtriangleup $ denotes the symmetric difference of sets.
Of course, by the Law of Large Numbers $\vert A_{N , \go}\vert /N 
\stackrel{N \to \infty}{\longrightarrow} 1/2$, $\bbP (\dd \go)$-a.s..
Therefore for the partition function  restricted to $\gO_{N, \go, \gep }$ we have
\begin{equation}
Z_{N, \go}^{ \gb, 0} \left(\gO_{N, \go, \gep }^\complement \right) \, \le\, 
\exp\left(  \gb  \vert A_{N , \go}\vert  (1-\gep) \right)
\, \stackrel{N \ge N_0(\go)}{\le}\,
\exp\left( N\frac \gb 2 \left(1-\frac \gep 2 \right)\right),
\end{equation}
for some $N_0(\go)$ which is  $\bbP (\dd \go)$-a.s. finite.
We can therefore focus on 
$Z_{N, \go}^{ \gb, 0} \left(\gO_{N, \go, \gep } \right)$,
which is bounded above by
$\exp (\gb \vert A_{N, \go}\vert) \bP \left(\gO_{N, \go, \gep }  \right)$
and  it is thus sufficient to show that
\begin{equation}
\label{eq:toconclude}
\limsup_{\gep \searrow 0}\limsup_{N\to \infty}
\frac 1N \log  \bP \left(\gO_{N, \go, \gep }\right) \le Q,
\end{equation}
to conclude.

In order to establish \eqref{eq:toconclude} we introduce a {\sl coarse
  graining} length $L\in \N$ ($L$ is sent to $\infty$ in the end, that
is after $N \to \infty$ and $\gep \searrow 0$, so that, in particular,
$\gep L $ can be chosen arbitrarily small and we assume below that
$\gep L \ll 1$). We assume that $N/L\in \N$ and we break $\{1, \ldots
, N\}$ into $N/L$ non-overlapping blocks $\{B_j\}_{j=1, \ldots, N/L}$
of length $L$.  For every realization of the disorder $\go$ we
decompose the event $\gO_{N, \go, \gep}$ into the disjoint union of
the events $\gO_{N, \go, \gep, \underline{v}}$, $\underline{v} \in
\{0,1\}^{N/L}$, defined by the property that if $S\in \gO_{N, \go,
  \gep, \underline{v}}$ then $B_j \cap \{n: \, S_n=0\}= B_j\cap A_{N,
  \go}$ if and only if $v_j=0$.  In short, there is a mismatch in the
block $B_j$ with respect to the energetically optimal contact
configuration if and only if $v_j=1$ (the mismatch can be on a single
site or on several sites).  Note that there cannot be more than $\gep
N$ blocks containing a mismatch, that is  if $\vert
\underline{v}\vert:=|\{i:v_i=1\}| > \gep N$  the event $\gO_{N,
  \go, \gep, \underline{v}}$ is empty.  Since $\gep L \ll 1$, only a
small fraction of the blocks contains mismatches.  We have therefore
\begin{equation}
\bP \left(\gO_{N, \go, \gep }\right) \,= \,
\sum_{\vert \underline{v}\vert \le \gep N}
 \bP \left(\gO_{N, \go, \gep,  \underline{v}}\right)
 \, \le \, \exp\left( c(\gep L) N/L\right)
 \max_{\vert \underline{v}\vert \le \gep N}
 \bP \left(\gO_{N, \go, \gep,  \underline{v}}\right),
\end{equation}
  where $c(x) \stackrel{x\searrow 0} \longrightarrow 0$ comes
  from estimating the cardinality of $\{ \underline{v}:\, \vert 
  \underline{v}\vert \le \gep N\}$.
  
 We are left with estimating $ \bP \left(\gO_{N, \go, \gep,
 \underline{v}}\right)$ uniformly in $ \underline{v}$. For this we
 introduce the random variable $\kappa_j =\inf\{n\in
 B_j:\,
\go_n=1\}$ ($\kappa_j =\infty$ if $\go_n=0$ for every $n\in B_j$) and
 for every $j$ such that $\kappa_j < \infty$
 the event 
 \begin{equation}
 E^j_{L, \go } \, :=\, \left\{ S: \, B_j \cap \left\{
 n: \, S_n=0\right\}\cap[\kappa_j,N]= A_{N, \go}\cap B_j\cap[\kappa_j,N]
\right\}, 
 \end{equation} that is simply the event that there is no mismatch in
 $B_j$ from step $\kappa_j$ onward. If $\kappa_j =\infty$ then
 $E^j_{L, \go }$ is just the set of all possible polymer trajectories,
 without any restriction.  Finally, we define
\begin{equation}
Y_j(\go)\, :=\, \begin{cases}
\bP\left(E^j_{L, \go } \big\vert S_{\kappa_j}=0\right) &
\text{if } \kappa_j < \infty, \\
1 & \text{otherwise. }
\end{cases}
\end{equation}
Note that $\{Y_j\}_j$ is a sequence of IID random variables and that,
by the Renewal Theorem (very much like in the proof of the lower
bound:  we are essentially evaluating the same quantity),
$(1/L) \log Y_1 (\go) \stackrel{L\to \infty}\longrightarrow Q$,
$\bbP(\dd \go)$-a.s.. Therefore, for every $\gd\in (0,1)$ we can find $L_0$
such that for $L\ge L_0$
\begin{equation}
\label{eq:first}
\bbP \left(Y_1(\go) \, \le \,   \exp((Q+\gd)L) \right)\, \ge \, 1- \frac{\gd}2.
\end{equation}
Notice now that 
\begin{equation}
\label{eq:second}
\bP \left(\gO_{N, \go, \gep,  \underline{v}}\right) \,  \le \, 
 \bP \left( \cap_{j:\, v_j=0} E^j_{L, \go }
 \right),
\end{equation}
and if we set  $k:=\vert\{i:\, v_i =0\} \vert $ and $\{i:\, v_i =0\}=\{i_1, \ldots, i_k\}$
($\{i_l\}_{l=1, \ldots, k}$  increasing)
we have that if $\kappa_{i_k}<\infty$
\begin{equation}
\label{eq:Yj}
 \bP \left( \cap_{j:\, v_j=0} E^j_{L, \go }
 \right)\, =\,  \bP \left( \cap_{l=1}^{k-1} E^{i_l}_{L, \go }
 \cap \left\{ S_{\kappa_{i_k}}=0\right\} \cap E^{i_k}_{L, \go }
 \right) \, \le \, \bP \left( \cap_{l=1}^{k-1} E^{i_l}_{L, \go }\right) Y_{i_k}(\go),
\end{equation}
where we have used the Markov property of $S$ and in the last step
we have neglected the event $\{ S_{\kappa_{i_k}}=0\}$.
With our definition of $Y_j (\go)$, the factorization inequality in
\eqref{eq:Yj} actually holds also for 
$\kappa_{i_k}=\infty$, so that by iterating we obtain
\begin{equation}
\label{eq:third}
 \bP \left( \cap_{j:\, v_j=0} E^j_{L, \go }
 \right)
 \, \le \,
 \prod_{j:\, v_j=0} Y_j (\go).
\end{equation}
By putting equations \eqref{eq:first} to \eqref{eq:third}
together, by applying the Strong Law of Large
Numbers and by exploiting the fact that more than a fraction $(1-\gep L)$ of the
$N/L$ blocks is free of mismatches,
we obtain that 
\begin{equation}
 \bP \left(\gO_{N, \go, \gep,  \underline{v}}\right) 
 \, \le \,
  \exp\left( (1-\gd)(Q+\gd)L (1- 2\gep L) \frac N L \right),
\end{equation}
 for $N\ge N_0(\go)$, with
$N_0(\go)$ a random value that is $\bbP(\dd \go)$-a.s.  finite. 
\smallskip

Overall we have therefore established that $\bbP (\dd \go)$-a.s.
\begin{equation}
\limsup_{N \to \infty} \frac 1N \log \bP \left( \gO_{N, \go, \gep }
\right) \, \le \, \frac{c(\gep L)}L + (1-\gd) (Q+\gd) (1-2\gep L) \stackrel{\gep\searrow 0}
\longrightarrow (1-\gd)(Q+\gd).
\end{equation}
Since $\gd$ can be chosen arbitrarily small the proof of the upper
bound is complete. 
\qed

\subsection{Estimates in the periodic case}
\label{sec:AppT}

We prove here Theorem~\ref{th:Tresults}.
The first result, that is the small temperature expansion 
in \eqref{eq:Tres1}, just comes from evaluating the ground
state energy. We omit the details since they are substantially
easier than the ones needed for the analogous result in 
the disordered set-up ({\sl cf.} Lemma~\ref{th:19} and App.~\ref{sec:A1}).
We point out also that such a result follows directly from the semi-explicit solution
available for periodic models \cite{cf:BG,cf:CGZ3} that we now outline since
we exploit it in order to establish the high temperature expansion
\eqref{eq:Tres2}.

The free energy of periodic models 
can be expressed by first introducing the (Abelian) group $\bbS:=
\Z/ \left(\tT (\go)\Z\right)$, that is $\{1, \ldots, \tT (\go)\}$ with
periodic boundaries. With abuse of notation 
an element $\ga$ of $\bbS$ is going to be identified
with a point in $\{1, \ldots, \tT(\go)\}$, 
 so by $n\in \ga$
we mean $n =k \tT(\go)+\ga$ for some $k \in \Z$.
For $b\ge 0$ we set 
\begin{equation}
K_\ga (b)\, :=\, \sum_{n \in \ga } K(n) \exp(-bn),
\end{equation}
and in turn for $\ga$ and $\gamma \in \bbS$
\begin{equation}
A_{\gamma, \ga} (b, \gb)\, :=\, K_{\ga -\gamma} (b) \exp \left( \gb \go_{\ga}\right).
\end{equation}
By the Perron--Frobenius theory on 
matrices with positive entries, the $\tT (\go) \times \tT (\go)$--matrix
$A(b, \gb)$ has a maximal positive eigenvalue, often called {\sl spectral radius}
of   $A(b, \gb)$, that we denote by $\gl_\go(b, \gb)$. By standard
arguments one shows that $\gl_\go(\cdot, \gb)$ is decreasing and smooth. 
In \cite{cf:BG, cf:CGZ3} it is shown that the  ($\go$--dependent) free energy 
 is $0$ if $\gl_\go(0, \gb)\le 1$.
  If instead $\gl_\go(0, \gb) >1$ then there exists a unique solution
 $b>0$ to the equation $\gl_\go(b, \gb)=1$ and such a $b$ is precisely
 the free energy $\tf(\gb)$.
 
 Let us now expand $A(b, \gb)$ for small values of $b$ and $\gb$:
 \begin{equation}
 A_{\gamma, \ga}(b, \gb)\, =\, 
 A_{\gamma, \ga}(0,0) 
 \left(1+ \gb \go_\ga +\frac 12 \gb^2 +
 o(\gb^2) \right) - \frac{\sqrt{2p}}{\tT(\go)}
 \left(b^{1/2} +o( b^{1/2}) \right),
 \end{equation}
 where the last term follows from the Riemann sum approximation
 procedure
 \begin{equation}
 \begin{split}
 \sum_{n \in \ga} \left( 1 - \exp(-bn) \right) K(n)\, &=
\, b^{1/2} \left( \sqrt{p/(2\pi)} +o(1)\right) b \sum_{n \in \ga}
 \frac{\left( 1 - \exp(-bn) \right)}{(bn)^{3/2}}
 \\
  &=\, \frac 1 {\tT (\go)}
  b^{1/2} \left( \sqrt{p/(2\pi)} +o(1)\right) 
  \int_0^{\infty} \frac{(1-\exp(z))}{z^{3/2}}\dd z,
 \end{split}
 \end{equation}
 and by the fact that the integral in the last term is equal
 to $2\sqrt{\pi}$.
 We now use the fact that the maximal eigenvalue 
 $\gl (A+\gep B)$, $A$ matrix with positive terms and
 $\gep$ small, can be written up to $O(\gep^2)$
 terms as $\gl(A)+ \gep u\cdot B v$,
 with $u$ and $v$ respectively right and left eigenvectors
 of $A$ with eigenvalue $\gl(A)$,  normalized
 by setting $\sum_\ga v_\ga =1$ and $\sum_\ga u_\ga v_\ga =1$.
  In our case $A=A(0, 0)$ turns out to be bi-stochastic, so
  $\gl (A)$, before denoted  $\gl _\go(0,0)$, is equal to $1$
  and $v_\ga =1/\tT (\go)$, as well as $u_\ga=1$
  for every $\ga$.
  This leads to the expansion
  \begin{equation}
  \label{eq:dainome}
  \gl (b, \gb)\, =\, 1 + \gb \left( \frac 1{\tT (\go)} \sum_{n=1}^{\tT(\go)}
  \go_n \right) + \frac 12 \gb ^2 - \sqrt{2p} \, b^{1/2}+ r(b, \gb),
  \end{equation}
  where $ r(b, \gb)=
   O(\gb^2)+o(\sqrt{b})$ if   $\sum_{n=1}^{\tT(\go)}
  \go_n \neq 0$ and $ r(b, \gb)=
   o(\gb^2)+o(\sqrt{b})$ otherwise.
  Therefore the existence of a (unique) solution $b=\tf(\gb)$ to
    $\gl _\go (b, \gb)=1$ for $\gb$ small requires
    $\sum_{n=1}^{\tT(\go)}
  \go_n \ge 0$ and 
  \begin{equation}
  \tf (\gb) \, =\, \begin{cases}
   \frac{\gb^2}{2p} \frac 1{\tT(\go)}  \sum_{n=1}^{\tT(\go)}
  \go_n +o(\gb^2)  & \text{if } \sum_{n=1}^{\tT(\go)}
  \go_n>0,
  \\
  \frac{\gb^4}{8p} + o(\gb^4)
  & \text{if } \sum_{n=1}^{\tT(\go)}
  \go_n=0.
  \end{cases}
  \end{equation}
 Armed with these asymptotic behaviors,
 \eqref{eq:Tres2} follows from $\tg^{-1}(y)= \sqrt{2y(1+o(1))/p}$
 ($y\searrow 0$). 
  
  What happens when
  $\sum_{n=1}^{\tT(\go)}
  \go_n<0$ is that $\gl_\go (0, \gb)$ is smaller than $1$ for small $\gb$
  (see \eqref{eq:dainome}).
  And if we set $\gb_0:= \sup\{ \gb:\, \gl_\go (0, \gb)<0\}$
(note that $\gb< \infty$ unless $\go_n=-1$ for every $n$,
as one can see from
the large $\gb$ expansion) then one readily sees that
$f_c(\gb)>0$ for $\gb>\gb_0$ and $f_c(\gb)=0$
otherwise.
\qed

\section*{Acknowledgments}  We would like to thank Stu Whittington for interesting
conversations on the re-entrance phenomenon, and Thierry Bodineau for
his suggestions  on the proof of Proposition \ref{th:disord}.  This research
has been conducted in the framework of the GIP-ANR project JC05\_42461
({\sl POLINTBIO}).


\begin{thebibliography}{99}

\bibitem{cf:A} K. S. Alexander, 
 \textit{The effect of disorder on polymer depinning transitions},
math.PR/0610008

\bibitem{cf:A2}
K. S. Alexander,
{\sl Ivy on the Ceiling:  First-Order Polymer Depinning Transitions with
Quenched Disorder}, preprint.

 
 \bibitem{cf:BG}
 E.~Bolthausen and G.~Giacomin, \textit{Periodic copolymers at selective interfaces: a large deviations approach}, Ann. Appl. Probab. {\bf 15} (2005),  963--983.
 
 \bibitem{cf:CGZ3}  
F. Caravenna,  G.   Giacomin and   L. Zambotti, 
\textit{Infinite volume limits of 
polymer chains with periodic charges},
  arXiv.org e-Print archive: math.PR 0604426
  
  \bibitem{cf:Danilo}
C. Danilowicz, Y. Kafri, R. S. Conroy, V. W. Coljee, J. Weeks and M. Prentiss,  
\textit{Measurement of the Phase Diagram of DNA Unzipping in the 
   Temperature-Force Plane}, Phys. Rev. Lett. {\bf 93}(2004), 078101. 

\bibitem{cf:Feller1}  
W.~Feller, \textit{An introduction to probability theory and its applications}, Vol. I,  
 Third edition, John Wiley \& Sons, Inc.,   
New York--London--Sydney, 1968. 

\bibitem{cf:Kafri}
 Y.~Kafri and A.~Polkovnikov,
 \textit{DNA unzipping and the unbinding of directed polymers in a random media},
 cond-mat/0605173
 


\bibitem{cf:RP} G. Giacomin, \textit{Random polymer models}, Imperial College Press, in press
 (2006).

\bibitem{cf:GTdeloc} G.~Giacomin and  F.~L.~Toninelli, \textit{Estimates on path delocalization for copolymers
at selective interfaces}, Probab. Theor. Rel. Fields {\bf 133} (2005), 464-482.

\bibitem{cf:GTloc}
G.~Giacomin and  F.~L.~Toninelli, \textit{The localized
phase of disordered copolymers with adsorption},
ALEA~{\bf 1} (2006), 149--180.

\bibitem{cf:Iliev_force}  
G.~Iliev, E.~Orlandini, and S.G.~Whittington, 
 \textit{Adsorption and 
 localization of random copolymers subject to a force: the Morita 
 approximation},  Eur. Phys. J. B {\bf 40} (2004), 
 63--71.

\bibitem{cf:2d} N.C. Jain, W.E. Pruitt, {\sl The Range of Rando
Walk}, in {\sl Proceedings of the Sixth Berkeley Simposium on Mathematical
Statistics and Probability} (Univ. California, Berkeley, Calif., 1970/71),
Vol. III: Probability Theory, pp. 31-50, Univ. California Press, Berkeley,
Cakuf., 1972.

\bibitem{cf:LubNel}
 D.K.~Lubensky and  D.R.~Nelson,  
Pulling pinned polymers and unzipping DNA,
Phys. Rev. Lett. {\bf 85} (2000), 1572--1575.


\bibitem{cf:Marenduzzo}
D.~Marenduzzo,  A.~Trovato and A.~Maritan,  
\textit{Phase diagram of force-induced DNA unzipping 
in exactly solvable models},
 Phys. Rev. E {\bf 64} (2001), 031901 (12 pages).
 
 \bibitem{cf:OTW04}
 E. Orlandini, M.C.~Tesi and S.G~Whittington,  
\textit{Adsorption of a directed 
polymer subject to an elongational force}, 
 J. Phys. A: Math. Gen. {\bf 37} (2004), 1535--1543.

\bibitem{cf:Nthese} 
N.~P\'etr\'elis, 
\textit{Localisation d'un polym\`ere en interaction avec une 
interface},  Ph.D. Thesis (2006), Univ. de Rouen (F).


\end{thebibliography}
\end{document}